\documentclass[nobibnotes,longbibliography]{ws-ijmpe}
\usepackage[compress]{cite}
\usepackage{hyperref}
\usepackage{academicons}
\usepackage{url}
\usepackage{tabularx}
\usepackage{booktabs} 
\newcommand{\orcida}[1]{\href{https://orcid.org/#1}{\includegraphics[width=8pt]{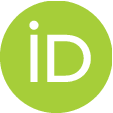}}}
\newcommand{\orcid}[1]{\href{https://orcid.org/#1}{\includegraphics[width=8pt]{orcid}} \href{https://orcid.org/#1}{https://orcid.org/#1}}
\renewcommand{\draftnote}{}

\begin{document}

\markboth{V. Kovalenko}{Evolution and fluctuations of chiral chemical potential in heavy-ion collisions}

\title{Evolution and fluctuations of chiral chemical potential in heavy ion collisions}

\author{Vladimir Kovalenko \orcida{0000-0001-6012-6615}\footnote{Corresponding author.}}

\address{
Saint Petersburg State University, \\ 7/9 Universitetskaya Nab., St. Petersburg 199034, Russia \\
v.kovalenko@spbu.ru}

\maketitle


\begin{abstract}
The possible appearance of the effects of local parity breaking in the QCD medium formed in heavy ion collisions due to violation of chiral symmetry can be quantified by corresponding chiral chemical potential $\mu_5$. The experimental observables sensitive to the effects of local parity violation in strong interaction include search for polarisation splitting of the $\rho_0$ and $\omega_0$ mesons via angular dependence of spectral functions in their decay to leptons. In this paper we study the space-time evolution and fluctuations of $\mu_5$ using relativistic hydrodynamics and estimate their effect on the light vector meson polarization splitting in Pb-Pb collisions at LHC energy.
\end{abstract}

\keywords{local parity violation; chiral imbalance; chiral chemical potential; relativistic hydrodynamics}

\ccode{PACS numbers: 11.30.Er, 11.30.Rd, 24.10.Nz, 13.20.-v}


\section{Introduction}


The strong interaction, as it is well known, obey the global spatial (P) parity conservation. So far, no evidence has been found for violation of P- and CP-symmetries in strong interactions.
However, quantum chromodynamics (QCD) does not forbid the local parity symmetry breaking due to large topological fluctuations at high temperature with dynamic generation of nontrivial topological charge configurations. A necessary condition for observing these effects is a sufficiently large space dimension and a long lifetime of a hot drop of QCD medium, which is available in central nuclear-nuclear collisions at the LHC \cite{KharzeevZhitnitsky,Buckley:1999mv,PhysRevD.70.074018}.

In the presence of a hot medium in the state of deconfinement, formed in heavy ion collisions,  the modification of the properties of hadrons is possible. They can arise due to the inhomogeneity of the medium, as well as due to such non-pertrubative effects as, for example, instantons \cite{BELAVIN197585}. 
In the perturbative calculations, the amplitudes of transitions between different degenerate vacua, connected by by topologically non-trivial gauge transformations, are equal to zero in any order of the perturbation theory. However, they can  afford for nonperturbative phenomena, known as sphaleron transition \cite{PhysRevD.43.2027,PhysRevD.61.105008,PhysRevD.67.014006}, happening through a potential barrier separating topologically nonequivalent vacua. Such configurations can be excited in the collision of heavy ions, or initially exist in them \cite{KharzeevZhitnitsky,Buckley:1999mv}.

So it can be assumed that parity-violating bubbles can arise in a collision of relativistic nuclei in a finite volume, causing the effect of a local parity nonconservation. The behavior of the topological charge of the gauge fields existing in a fireball for a long time, in statistical approach, can be described in terms of topological chemical potential ($\mu_\theta$).
Due to the partial nonconservation of the axial current in QCD, in the approximation of low fermion masses, it is possible to relate the topological charge to the value of the chiral imbalance, which is defined as the average difference between the number of right and left quarks in a fireball after heavy ion collisions at high energy \cite{PhysRevLett.81.512}.
Thus, chiral imbalance can lead to the formation of local parity violation (LPB) in a quark-hadron medium with local thermodynamic equilibrium characterized by an axial chemical potential \cite{Kharzeev:2004ey}. In this case, a connection arises between the topological ($\mu_\theta$) and axial ($\mu_5$) chemical potentials:

\begin{equation}
\mu_5 = \dfrac{1}{N_f}\mu_\theta,
\end{equation}
where $N_f$ is a number of flavors.

The effect of local parity nonconservation in strong interactions can be experimentally searched for using local observables sensitive to the processes at hadron space scale, which are not canceled out after the summation over the whole medium.

It was shown in \cite{ANDRIANOV2012230,Andrianov:2010eg,PlanellsAndrianovPoS,PhysRevD.90.034024} that this effect can be verified experimentally by analyzing the yields of dilepton pairs in the region of small invariant masses in heavy ion collisions. This will require simultaneous scanning of both the invariant mass ($m_{ll}$) and the expansion angle ($\theta_A$) of leptons from the decays of light vector mesons. 
In the case of a nonzero axial chemical potential, a polarization splitting of the spectral functions of $\rho$ and $\omega$ mesons occurs in a part of the phase space, with the formation of a characteristic two-peak structure.

Later, it was shown \cite{Kovalenko_2020} that if the radiative corrections are taken into account, then the presence of the axial chemical potential leads not only to the splitting of the masses of the left and right polarizations of vector mesons, but also to their general shift towards an increase. These effects depend both on the value of the axial chemical potential $\mu_5$ and on the momentum of the vector meson ($k$).

Because the polarization mass splitting depends on the axial chemical potential $\mu_5$, the experimentally measured picture would depend
also on the fluctuations of $\mu_5$. There is a risk that in case of sufficiently large fluctuations the splitting would smear out and become hard to detect. So the main purpose of the paper is to investigate the evolution of 
the axial charge density during heavy ion collision from the initial stages to freeze-out and to estimate its fluctuations induced by this evolution.


We should add that other ways of observation of parity nonconservation effects  in heavy ion collisions are possible.
In the presence of the large magnetic field, which is characterized by semi-central and peripheral ion collisions, the so-called chiral magnetic effect (CME) can happen \cite{Kharzeev:2004ey,KHARZEEV2008227}. It was measured at RHIC and LHC \cite{Wang_2017,HAQUE2019543, Aziz:2020nia} and a CME-like signal was found. However, other backgrounds, like a local charge conservation, play a comparable role \cite{Huang_2016}.
Nevertheless, the comparison of the experimental results with modeling \cite{Yuan:2023skl} showed the best agreement is achieved for the $\mu_5$ above 300 MeV. 

The evolution of the axial charge in the QCD medium is modeled using Anomalous-Viscous Fluid Dynamics (AVFD) framework \cite{Shi:2017cpu} for studying the signatures of chiral magnetic effect in relativistic nuclear collisions \cite{Shi:2019wzi, Jiang:2016wve} (see also review papers \cite{Kharzeev:2020jxw,Choudhury:2021jwd,Li:2020dwr}). The main attention is devoted to the calculation of chiral magnetic effect in presence of magnetic field in non-central heavy-ion collisions. In the present work we focus on modeling parity-violating effects \cite{ANDRIANOV2012230,Andrianov:2010eg,PlanellsAndrianovPoS,PhysRevD.90.034024} relevant also for central nucleus-nucleus collisions.

We should note that apart from the vector meson parity mass splitting \cite{ANDRIANOV2012230,Andrianov:2010eg,PlanellsAndrianovPoS,PhysRevD.90.034024} there are other proposals, also suitable for central heavy-ion collisions, including the 
search for decays of a scalar charged $a_0$ meson into a photon and a charged pion or into three charged pions\cite{Andrianov:2017meh,Andrianov:2017hbf,Andrianov:2018gim,AndrianovParticles2020, KovalenkoPetrovBull,KovalenkoPetrovPEPAN}, study the relative rate of
$\pi$ meson decay by muon-neutrino and electron-neutrino channels  \cite{AndrianovParticles2020}, search for possible asymmetry of the photon polarization \cite{refId0}.


\section{Methods}


The simulation of the evolution of the QCD medium produced in heavy ion collisions at LHC energy has been performed using relativistic hydrodynamics. A commonly-used package MUSIC \cite{MUSIC1,MUSIC2,MUSIC3,MUSIC4}, 3+1D relativistic second-order viscous hydrodynamics for heavy ion collisions, was used. In this studies, for simplicity, boost-invariant 2+1D version was taken. Lattice hotQCD equation of state (EOS) was used, and no dependence in EOS on $\mu_5$ were assumed. The viscosity over entropy density parameter is set as $\eta/s=0.08$.

The initial energy density distributions were generated using the Glauber Monte Carlo sampler \cite{MCGlauber,Shen:2014vra} with the initial entropy density deposited at wounded nucleon positions. Central Pb-Pb collisions at 5.02 TeV (with impact parameter $0<b<2$ fm) were considered.

\begin{figure}[h]
\centering
\includegraphics[width=13.5cm]{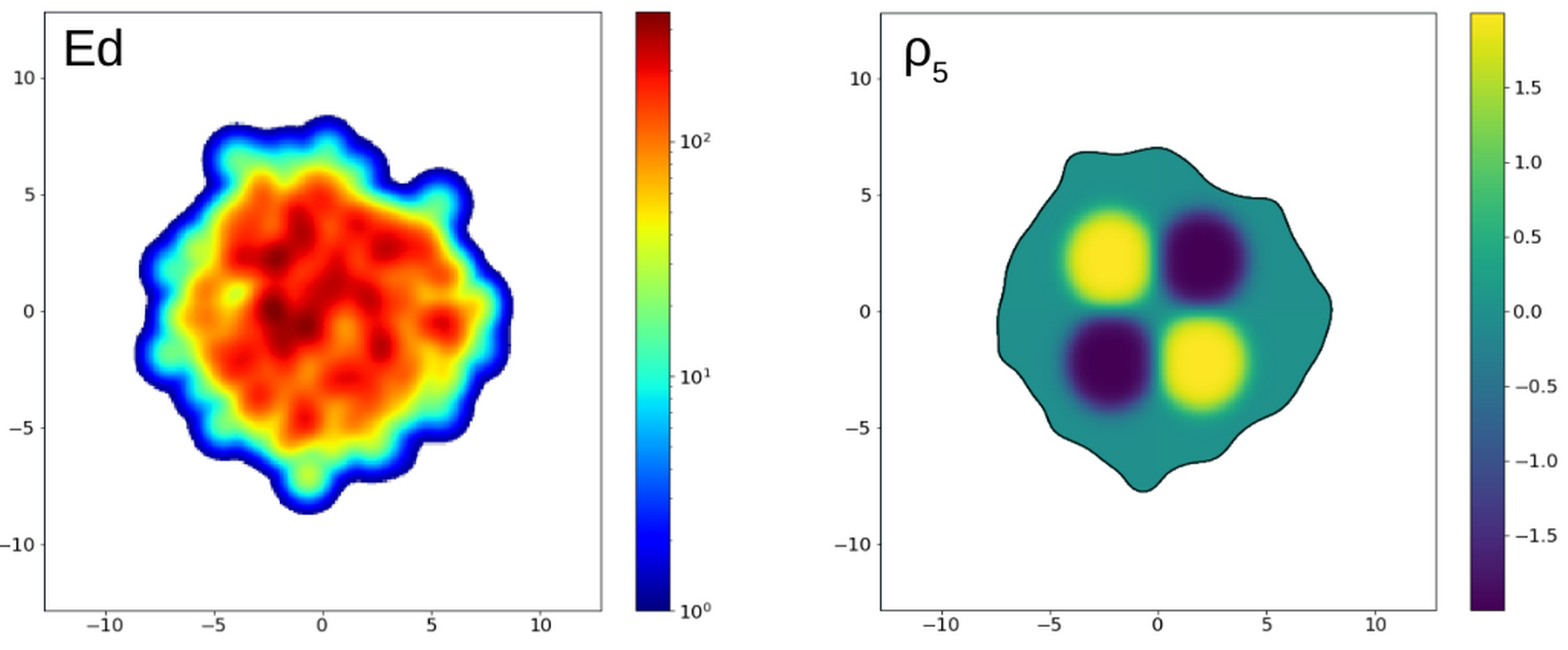}
\\ { \ \ \  \ \ \ \small(\textbf{a}) \ \ \ \ \ \ \ \ \ \ \ \ \ \ \ \ \ \ \ \ \ \ \ \ \ \ \ \ \ \ \ \ \ \ \ \ \ \ \ \ \ \ \ \ \ \ \ \ \ \ \ \ \ \ \ \ \  \ \ \ \ \ (\textbf{b})\ \ \ \ \ \ }
\caption{The initial distribution of energy density (\textbf{a}) and axial charge density (\textbf{b}) in the transverse plane as an example for one Pb-Pb collision event at an energy of 5.02 TeV.
\vspace{0.5cm}
\label{fig1}}
\end{figure}

The initial axial charge \cite{Lee:1974ma,Kharzeev:2007tn} densities were set up  as negative and positive isolated areas in the transverse plane with radius of 2.1~fm with the constant $\rho_5$ inside them and zero total axial charge  (see Figure~\ref{fig1}).
In the case of ideal evolution, the net axial current evolves according to the following equation \cite{Huang:2021bhj}:
\begin{equation}\label{mainequation}
\partial_\mu J_{5}^\mu=-\frac{N_c Q_f^2}{2 \pi^2} E \cdot B.
\end{equation}
In case of central collisions the magnetic field is small, so we put electromagnetic fields $B, E$ as zero. With this assumption the equation (\ref{mainequation}) becomes the continuity equation.

In order to take into account the viscous dissipative effects and the  damping of axial charge due to gluon topological fluctuations \cite{Huang:2021bhj}, the Anomalous-Viscous Fluid Dynamics (AVFD) framework \cite{Shi:2017cpu,AVFDgithub} was applied. In this framework it is possible to consider the 2+1D  hydrodynamic history of the QCD fluid, modeled with MUSIC \cite{MUSIC1,MUSIC2,MUSIC3,MUSIC4}, as a background and to perform the 2nd order evolution of the left-handed ($\chi=-1$ and right-handed $\chi=1$) fermion currents:

\begin{equation}
\begin{aligned}
\hat{D}_\mu J_{\chi, f}^\mu & =\chi \frac{N_c Q_f^2}{4 \pi^2} E_\mu B^\mu \\
J_{\chi, f}^\mu & =n_{\chi, f} u^\mu+\nu_{\chi, f}^\mu+\chi \frac{N_c Q_f}{4 \pi^2} \mu_{\chi, f} B^\mu \\
\Delta^\mu \hat{d}\left(\nu_{\chi, f}^\nu\right) & =-\frac{1}{\tau_r}\left[\left(\nu_{\chi, f}^\mu\right)-\left(\nu_{\chi, f}^\mu\right)_{N S}\right] \\
\left(\nu_{\chi, f}^\mu\right)_{N S} & =\frac{\sigma}{2} T \Delta^{\mu \nu} \partial_\nu\left(\frac{\mu_{\chi, f}}{T}\right)+\frac{\sigma}{2} Q_f E^\mu
\end{aligned}
\end{equation}

The dissipative effect is guided by the diffusion coefficient $\sigma$ and relaxation time $\tau_r$. Again, we put magnetic field as zero.

During the hydrodynamic evolution, there exist random topological fluctuations of the gluon fields that can influence
the axial current evolution.
In order to take into account the possible  damping of axial charge due to gluon topological fluctuations \cite{Huang:2021bhj}, the equation of the net axial current acquires an additional term, which depends on the relaxation time $\tau_{cs}$, so the equation (\ref{mainequation}) reads:
 \begin{equation}\label{mainequation2}
\partial_\mu J_{5}^\mu= - \dfrac{ n_{5}}{\tau_{cs}},
\end{equation} 
where $n_5$ is the net axial density. The uncertainty in the $\tau_{cs}$ is quite large raging from tenths to hundreds fm  \cite{Huang:2021bhj}.

The evolution of the axial density is traced up to freeze-out, which is defined by the condition that the  energy density $\text{Ed=0.18~GeV/fm}^3$.
Because the effect of the vector meson polarization splitting equally depends on the positive and negative axial charge, at freeze-out the distribution of the absolute value of the axial charge density is obtained.


To calculate the di-lepton invariant mass spectra from the vector mesons emitted from a medium with chiral imbalance, a Monte Carlo model based on the Pythia 8 event generator was used (version 8.2) \cite{pythia8.2}
 with a built-in Angantyr collision model for relativistic nuclei \cite{angantyr}. For the simulations, collisions of lead nuclei at an energy of 5.02 TeV were used.
In order to increase the statistics of the di-lepton spectrum, the fraction of decays of $\rho$ and $\omega$ mesons through the di-electron and di-muon channels was increased up to 0.44. 

Decay products were considered in the following rapidity intervals: 
\begin{equation}\nonumber
-0.8<\eta<0.8 \text{ for electrons, }-3.6<\eta<-2.45\text{ for muons.}
\end{equation}
This corresponds to the range of the time projection camera and the internal tracking system of the ALICE experiment at the LHC in the central rapidity region and the muon system in the forward rapidity region \cite{TheALICECollaboration_2008}. 
At was initially proposed, to perform the angular analysis in the central rapidity region \cite{PhysRevD.90.034024} and ensure the equal treatment of both the di-muon and di-electron channels, for all muon tracks we applied a boost along the z axis by rapidity of 3.05 so that in the new reference frame they lie in the midrapidity. Only di-leptons coming from the decay of light vector mesons were kept for analysis, and background events were not taken into account.

The spectral functions of $\rho$ and $\omega$ mesons are shifted in presence of the non-zero chiral charge so that their masses depend on the $\mu_5$ and momentum $k$ as \cite{Kovalenko_2020}:
\begin{equation}
m^*_{\rho,\omega} = m_{\rho,\omega} + 0.23\, \mu_5^2 + 1.37\, \mu_5\, k + 2.54 \, \mu_5^2\,k^2
\end{equation}
(here $m^*_{\rho,\omega}, m_{\rho,\omega}, \mu_5, k$ are in GeV).

Then the leptons momentum is smeared in order to take into account the experimental resolution. The parameters were taken as for the ALICE experiment in the LHC Run 3 conditions (after the LS2 upgrade) \cite{AbelevTRRITS_2014,ALICE_UPGR_Garcia-Solis:2015gkn,CERN-LHCC-2015-001}.
The resulting standard deviation  of transverse momentum for electrons was taken as 1\% and for muons as 0.5\%.

Then the di-lepton distributions are calculated with some cuts on the angle $\theta_A$ between the leptons \cite{PhysRevD.90.034024}. The cuts are selected to keep the reasonable statistics but ensure as good as possible the separate the lepton polarisations.


\section{Results}

The initial distributions of energy density and axial charge density over the transverse plane are shown in Figure~\ref{fig1} for one example Pb-Pb event.

\begin{figure}[h]
\centering
\includegraphics[width=13.5cm]{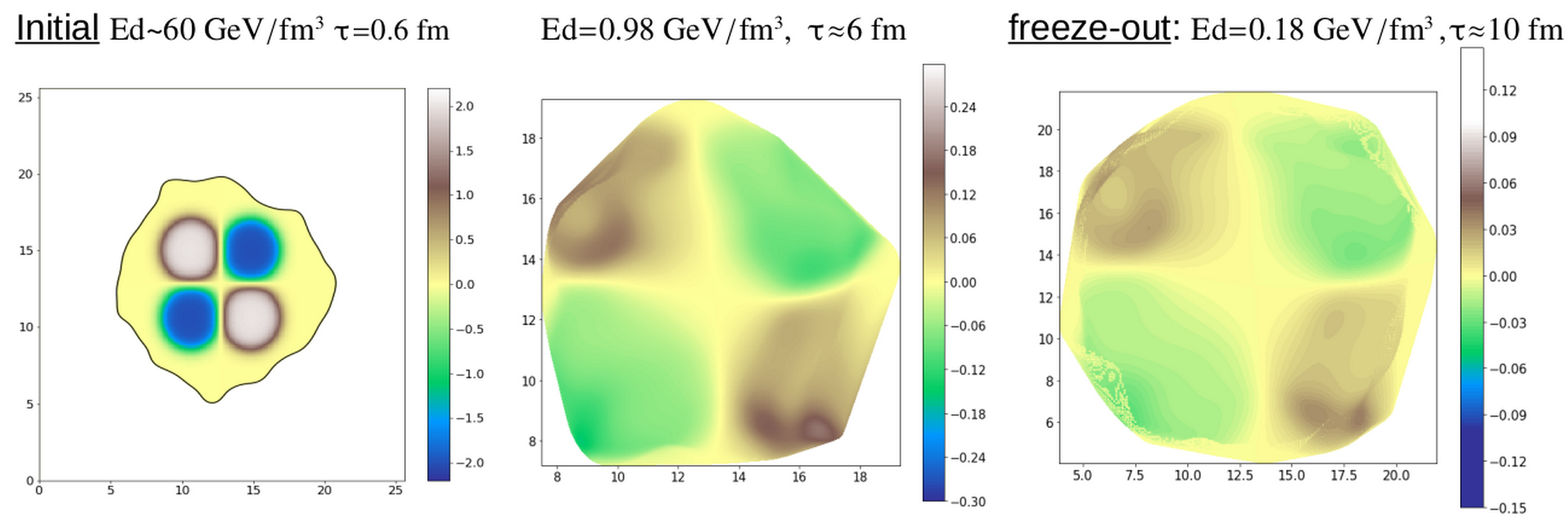}
\caption{Hydrodynamic evolution of the axial charge density from initial states to freeze-out, an example for one Pb-Pb collision event at an energy of 5.02 TeV.\label{fig2}}
\end{figure}

Then the medium undergo the hydrodynamic evolution. Some intermediate stages are shown in Figure~\ref{fig2}. 
On the 2nd and 3rd panels the snapshots of the axial charge density at the fixed energy density are plotted, which approximately correspond to the fixed proper time $\tau$. The results demonstrate that some smearing is happening and also the neutral area in the center is increasing, however overall the local charge excess stays throughout the entire evolution up to the freeze-out. 
Due to expansion, the overall scale of the charge density decreases.

\begin{figure}[h]
\includegraphics[width=7.5 cm]{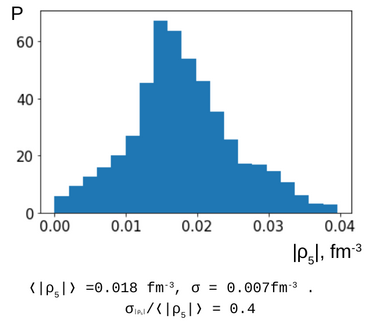}\centering
\caption{Distribution of the absolute value of the axial charge density at the freeze-out.\label{fig3}}
\end{figure}

Because the effects depend on the absolute value of the axial charge, in Figure~\ref{fig3} we show the distribution of the absolute value of the axial density at the moment of freeze-out,  and calculate the mean value $|\rho_5|$ and standard deviation $\sigma_{|\rho_5|}$ of this distribution.

Then the value of relative fluctuation $\delta_{|\rho_5|} = \sigma_{|\rho_5|} / \langle{|\rho_5|\rangle}$ is calculated.\
After the event-by-event study of this value  the following result was obtained: 
$\delta_{|\rho_5|}=0.42\pm0.04$.

\unskip

\begin{figure}[h]
\centering
\includegraphics[width=12.5cm]{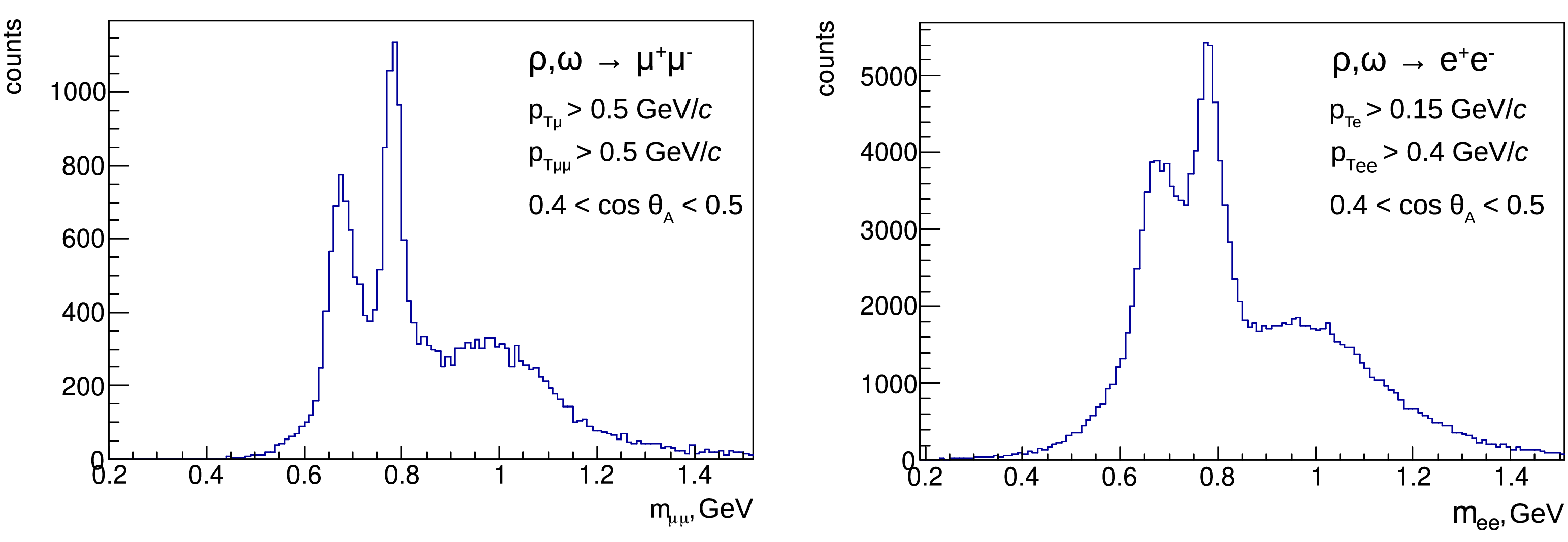}
\\ { \ \ \ \ \ \ \ \ \ \small(\textbf{a}) \ \ \ \ \ \ \ \ \ \ \ \ \ \ \ \ \ \ \ \ \ \ \ \ \ \ \ \ \ \ \ \ \ \ \ \ \ \  \ \ \ \ \ \ \ \ \ \ \ \ \ \ \ \ \ \ (\textbf{b})}
\caption{Invariant mass distribution of di-muons \textbf{(a)} and di-electrons \textbf{(b)} in the Monte Carlo model from the decays of $\rho$ and $\omega$ mesons under the expected conditions of the ALICE Run 3 experiment at \emph{$\mu$}\textsubscript{5} distributed according to Gauss with a mean of 0.15~GeV and a standard deviation $\sigma_{\mu_5}$ = 0.06~GeV ($\delta$\textsubscript{\emph{$\mu$}}=40\%).\label{fig4}}
\end{figure}

Taking into account the estimation of the axial charge density fluctuation we can include it in the calculation of the di-lepton mass distributions.
In case of the local thermodynamic equilibrium the  axial charge density and axial chemical potential are related,
and in the range of our consideration (a few hundreds MeV) the dependence is close to linear. So we put $\delta_{|\rho_5|} \approx \delta_{|\mu_5|}$.

In the Figure~\ref{fig4} the di-electron and di-muon mass distributions are plotted taken into account the fluctuation of the axial chemical potential at the level of 40\%, the experimental resolution of ALICE in the Run 3 conditions and the kinimatic cuts and selection with angle $\theta_A$ between leptons (also shown in the figure). The mean value of axial chemical potential was taken as $\mu_5=0.15$~GeV. The results show that the separation of at least two polarisations both for electrons and muons is visible in this conditions.

\section{Dissipative effects}

\begin{figure}[h]
\vspace{-0.2cm}
\centering
\includegraphics[width=10cm]{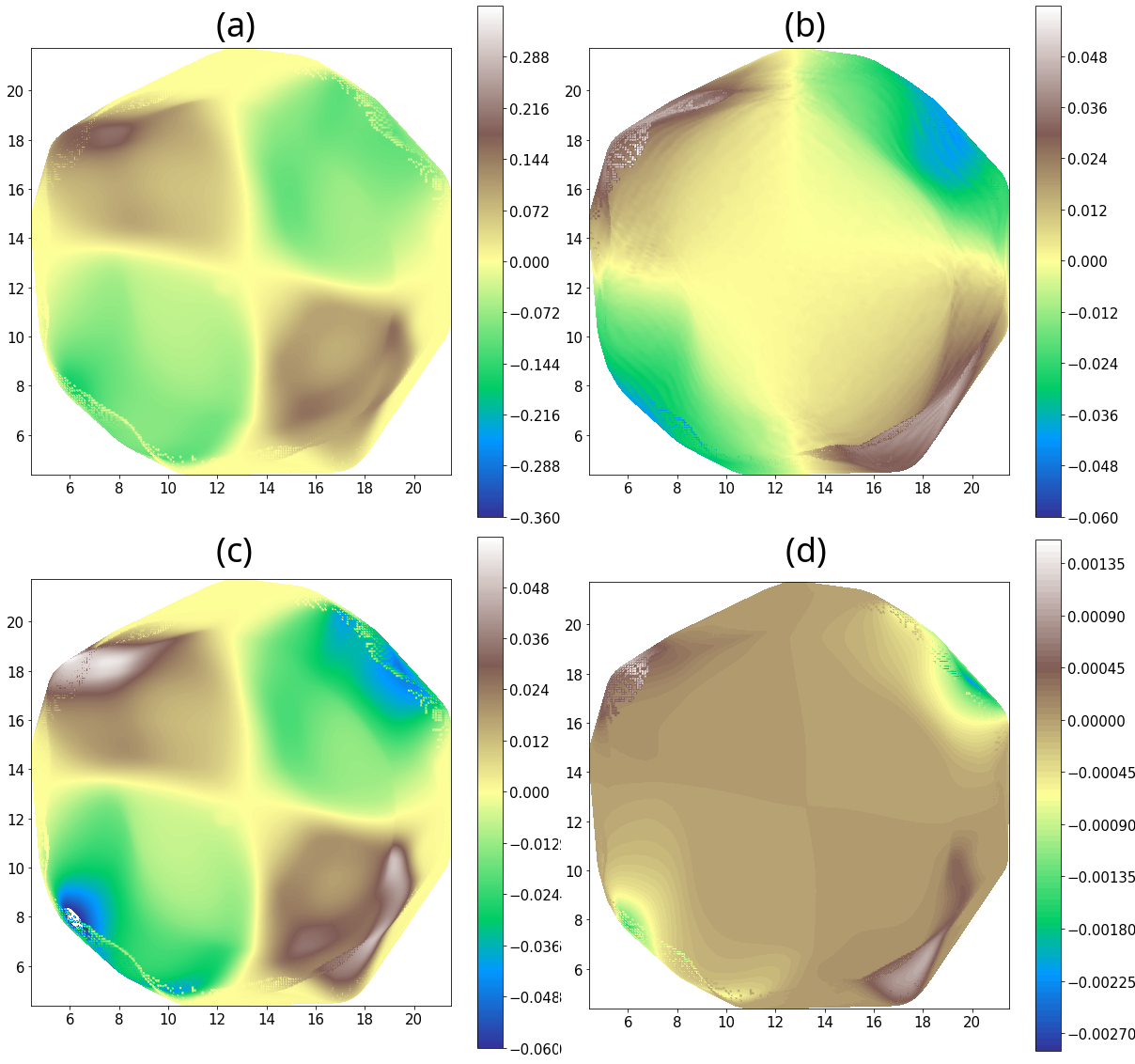}
\vspace{-0.2cm}
\caption{Modification of freeze-out axial charge distribution: (a) ideal evolution, (b) with viscous effects ($\sigma = 0.3 T$ and $\tau_r = 0.5/T$), (c) with axial charge damping ($\alpha_s=0.2$), (d) with axial charge damping ($\alpha_s=0.3$). \label{fig5}}
\end{figure}  

In figure \ref{fig5} we show for one event, how the results of the axial charge evolution up to freeze-out are modified due to the inclusion of normal viscous effects (diffusion coefficient $\sigma$ and relaxation time $\tau_r$. We use the values  \cite{Shi:2017cpu,Denicol:2012vq} $\sigma = 0.3 T$ and $\tau_r = 0.5/T$, where $T$ is the temperature.

We calculated the relative fluctuation $\delta_{|\rho_5|} = \sigma_{|\rho_5|} / \langle{|\rho_5|\rangle}$ and see that it is increased from 0.39 to 0.44 due to additional smearing caused by viscous effects. However, this increase is quite moderate so it does not affect much the possibility of observing the vector meson polarization splitting.

We also checked the effect of axial charge damping due to gluon topological fluctuations with parameters in the range from $\alpha_s$=0.1, $\tau_{cs}=113\rm{fm}/c$ to $\alpha_s$=0.3, $\tau_{cs}=1.4\rm{fm}/c$ \cite{Huang:2021bhj} (see table \ref{tab1}). 
We see that if the Chern-Simons relaxation time is higher than or comparable with the characteristic time of the QGP evolution (up to freeze-out), the distribution of the axial charge survives. However, for small values of $\tau_{cs}$ the medium is considerably smoothed out leaving a small chance to observe parity violation effects at the freeze-out.

\begin{table}[h] 
\vspace{-0.2cm}
\caption{Effect of axial charge damping due to gluon topological fluctuations.\label{tab1}}
\newcolumntype{C}{>{\centering\arraybackslash}X}
\newcolumntype{Y}{>{\arraybackslash}X}
    \centering
\begin{tabularx}{0.4\textwidth}{ccc}
\toprule
{$\alpha_s$}	& {$\tau_{cs}$ (at 300GeV)}	& {$\delta_{|\rho_5|}$}\\
\midrule
0.1		& 113 fm/$c$			& 0.39\\
0.2	& 7.1 fm/$c$			& 0.48 \\
0.3	& 1.4 fm/$c$			& 0.80 \\
\bottomrule
\end{tabularx}
\end{table}

\section{Conclusions}

The evolution of the axial chemical potential during the hydrodynamical expansion of the QCD medium, produced in Pb-Pb collisions at LHC energy, has been studied.
The results show that the regions of local axial charge excess survive in the medium throughout the entire evolution of the fireball, from the initial stages to freeze-out.
The resulting smearing induced by the hydrodynamical evolution is obtained at the level of 40\%. The viscous effects moderately increase this variance, leaving the axial charge splitting at a satisfactory level. The effect of gluon topological fluctuations strongly depends on the relaxation time and can smear out the axial charge distribution if $\tau_{cs}$ is below the freeze-out time.

The study of the influence of such fluctuation on the possibility of detecting local space parity non-conservation via angular analysis of di-lepton distribution showed that such fluctuations do not destroy the possibility of observing the splitting of vector meson polarizations inder the conditions of LHC Run 3 data.

The estimates done in this paper show that a more detailed modeling of initial conditions and the building of a complete framework
with parity-violating effects
 based on EBE-AVFD is deserving. It will allow to study the centrality-dependent effects (including magnetic field) and this can be done in further works.

\section*{Acknowledgements}
The study was funded by the Russian Science Foundation grant No. 22-22-00493, \url{https://rscf.ru/en/project/22-22-00493/}

\section*{ORCID}
Vladimir Kovalenko \orcid{0000-0001-6012-6615}
\bibliographystyle{woc}

\begin{sloppypar}
\setlength{\emergencystretch}{3em} 
\sloppy \bibliography{bibtexfile} \sloppy
\end{sloppypar}

\end{document}